\documentclass[12pt,BCOR0.1cm,DIV14]{scrartcl}
\usepackage[pdftex]{graphicx}
\usepackage{amssymb,amsfonts,amsmath}
\usepackage{xcolor}
\usepackage{caption}
\usepackage{subcaption}
\usepackage{tabularx}
\usepackage{float}
\newcommand{\fref}[1]{Fig. \ref{#1}}
\newcommand{\tref}[1]{Table \ref{#1}}
\newcommand{\Fref}[1]{Fig. \ref{#1}}
\newcommand{\eref}[1]{Eq. \ref{#1}}
    \makeatletter
    \let\@fnsymbol\@arabic
    \makeatother
\usepackage[round,numbers,sort&compress]{natbib} 

\definecolor{shadecolor}{RGB}{255,50,50}

\begin{document}

\title{Conformational mechanism for the stability of microtubule-kinetochore attachments}

\author{
Zsolt Bertalan\thanks{Institute for Scientific Interchange Foundation, Via Alassio 11/C, 10126 Torino, Italy}
\and Caterina A. M. La Porta\thanks{Department of Biosciences, University of Milano, via Celoria 26, 20133 Milano, Italy}
\and Helder Maiato\thanks{Chromosome Instability and Dynamics Laboratory, Instituto de Biologia Molecular e Celular, University of Porto, Porto, Portugal}~$^{,}$\thanks{Cell Division Unit, Department of Experimental Biology, Faculty of Medicine, University of Porto, Porto, Portugal}
\and Stefano Zapperi\footnotemark[1]~$^{,}$\thanks{National Research Council of Italy, Istituto per l'Energetica e le Interfasi, Via R. Cozzi 53, 20125 Milano, Italy}~$^{,}$\thanks{Corresponding author.}
}

\maketitle

\begin{abstract}Regulating the stability of microtubule(MT)-kinetochore attachments is fundamental to avoiding mitotic errors and ensure proper chromosome segregation during cell division. While biochemical factors involved in
this process have been identified, its mechanics still needs to be better understood. Here we introduce and
simulate a mechanical model of MT-kinetochore interactions in which the stability of the attachment is
ruled by the geometrical conformations of curling MT-protofilaments entangled in kinetochore fibrils.
The model allows us to reproduce with good accuracy in vitro experimental measurements of the detachment times
of yeast kinetochores from MTs under external pulling forces. Numerical simulations 
suggest that geometrical features of MT-protofilaments may play an important
role in the switch between stable and unstable attachments.


\emph{Abbreviations:}{ MT, microtubule; PF, protofilament}
\end{abstract}

\newpage

\section*{Introduction}

During mitosis, the cell equally divides into two daughter cells, each getting a copy of the original genetic material. Successful division requires that the two identical sister chromatids of mitotic chromosomes attach to the plus-ends of spindle microtubules (MTs) via their kinetochores \cite{cheeseman2008}. This process is critical 
because incorrect attachments lead to mitotic errors giving rise to genetic instabilities, also involved in cancer \cite{abbas2013}. 
To ensure accurate chromosome segregation, correct MT-kinetochore attachments should remain stable while faulty attachments should be destabilized and corrected  \cite{tanaka,biggins-murray,lampson-cheeseman}. 

MTs are composed by a number of protofilaments (PFs), typically thirteen, and polymerize by the addition of tubulin dimers in their GTP-bound state. Growing MTs switch to a shrinkage phase when all or most GTP-bound tubulin is hydrolysed, a process known as catastrophe, while the switch back to a growing state is known as resuce. These two processes constitute the {\it dynamic instability} of MTs.\cite{mitchison1984}. GTP-hydrolysis also induces a change in conformation of MT protofilaments from a straight to a curved state \cite{hyman1998,wang2005} which eventually leads to depolymerization, since curved PFs tend to peel from the MT while straight filaments are stable. Indeed electron micrographs of microtubules show that individual PFs can be seen curving outwards from the ends \cite{chretien1999}. Mechanical measurements of the rigidity of MTs  show that their Young's modulus is two orders of magnitude smaller than the shear modulus \cite{Kis}. This implies that tubulin dimers interact strongly along the PF and weakly along the transverse direction. All these structural, mechanical and kinetic aspects have been included in theoretical and computational models that describe with great accuracy the main features of the stability and dynamic instability of MTs \cite{janosi1998,janosi2002,vanburen2005,hunyadi2007,zapperi-mahadevan}.  

MT-kinetochore attachments vary in different organisms, but all seem to share a common feature \cite{mcintosh-fibrils, mcintosh-MTang}: Fibrils extending from the kinetochore either link directly to curling MT-PFs, as in most higher eukaryotes \cite{dong-ktmt,schmidt-ska1}, or fibrils linking to a ring, known as the $Dam$1 complex \cite{tanaka-yeast,volkov}, form around the attached MT, as in budding yeast \cite{westerman-damform,westerman-damprog}. Once the attachment has been formed,  MT depolymerization \cite{koshland1988} provides a force that is strong enough to carry kinetochore-attached loads \cite{grishchuk-yeast,grishchuk-MTs,tanaka-yeast}, even in the absence of motor proteins. The precise nature of the attachment between kinetochore fibrils and MTs has been the object of
intense experimental investigation suggesting that binding occurs through the Ndc80 complex which interacts with tubulin by weak electrostatic forces \cite{ciferri2008}.  Other experiments show that the Ndc80 complex acts like a curvature sensor and binds 
preferentially to straight MT PFs, typical of polymerizing MTs \cite{alushin2010}. On the other hand, electron micrographs illustrate vividly that kinetochore fibrils directly connect to the tips of {\it curling} PFs \cite{mcintosh-fibrils}. 

In a recent paper, Akiyoshi et al. \cite{akiyoshi} have shown that yeast kinetochores form {\it catch-bonds}  with MTs. {\it Catch-bonds} become stronger under a pulling force \cite{marshall-catchbond,thomas-catchbond}, thus providing a possible stabilizing mechanism needed for chromosome segregation. Akiyoshi et al. also show that the kinethocore forms {\it catch-bonds} only 
with depolymerizing MTs, whereas it forms standard force-weakening bonds with polymerizing MTs \cite{akiyoshi}.
The authors encompass this information into a simple two-states kinetic model that is able to fit well
the experimentally measured detachment times \cite{akiyoshi}. The physical mechanism by which a {\it catch-bond} 
forms remains, however, still unclear. It is easy to understand that electrostatic interactions alone, even if protein complexes act cooperatively \cite{ciferri2008,zaytsev2013}, would not give rise to a {\it catch-bond}: pulling charges apart leads to a weakening of the bond and should therefore increase the detachment rate. Thus electrostatic interactions reasonably account for the experimentally observed increase of detachment rate for polymerizing MTs under tension, but can not
explain at the same time the decreasing detachment rate observed for depolymerizing MTs. 

Models of MT-kinetochore interactions are abundant ranging from the classical sleeve model \cite{hill-sleeve,joghunt} to curling models \cite{mcintosh-ktmt,molodtsov1,efremov} and syntheses of both \cite{shtylla,shtylla2} as well as incorporation of motor proteins \cite{civek}, but none of them provide insight into the observed  {\it catch-bond} behavior. In this paper, we resolve this puzzle by combining existing experimental evidence into a model of 
MT-kinetochore attachment that can explain the formation of {\it catch-bonds}. We consider the interaction of a microtubule with a set
of kinetochore fibrils with tips that can directly bind to straight MT-protofilaments and to neighbor fibril tips as suggested in Ref. \cite{ciferri2008} for the Ndc80 complex. Similarly, the fibrils can be effectively cross linked by other protein complexes as e.g. the $Dam1$ complex in budding yeast  or the Mis12 complex in PtK$_1$ cells \cite{dong-ktmt}. Hence, when a polymerizing MT approaches the kinetochore, fibrils attach to its surface due their direct interactions. This attachment is a standard force-weakening bond as expected, but the mutual interactions or cross-linking between fibrils naturally leads to the formation of fibril loops. When MTs depolymerize, the direct fibril-MT binding force is strongly suppressed
\cite{alushin2010}, but the tips of curling PFs can easily entangle in the fibril loops. This attachment is now a {\it catch-bond}, since it
becomes stronger under tension. The mechanism we propose is very general and could also involve
other kinetochore proteins, such as CENP-E \cite{gudimchuk2013}, CENP-T \cite{suzuki2011}, CENP-F \cite{mcintosh-fibrils} or the Ska complex \cite{albaabad2014}, rather than just Ndc80, which is too short to account for the fibrils alone.
Finally, the entangled organization of the observed in vertebrate kinetochores \cite{dong-ktmt}, provides another striking example where a velcro-like attachment, such as the one we propose, could naturally take place.

We illustrate the formation of a conformational attachment by three dimensional simulations of a single 
depolymerizing MT interacting with a set of kinetochore fibrils. Once we are confident that a conformational
attachment is formed, we can reduce the computational complexity of the problem by focusing on a two-dimensional
representation of the interaction between a PF and a fibril loop. Numerical results of the two dimensional
model reproduce with good accuracy the {\it catch-bond} behavior reported experimentally in Ref. \cite{akiyoshi}. We also study the stability of the MT attachment and find that it crucially depends on the local conformation of the MT. By changing the intrinsic curvature of the MT-PF, the attachment is destabilized and the {\it catch-bond} disappears. Our results suggest that the experimentally observed tension-induced stabilization of MT-kinetochore attachments could be explained by a conformational mechanism although 
chemical affinities between MTs and kinetochore proteins may also play a role.

\section*{Materials and Methods}

\subsection*{Three dimensional model of MT-kinetochore interactions}

\begin{figure}[ht]
\centering
\includegraphics[width=0.7\textwidth]{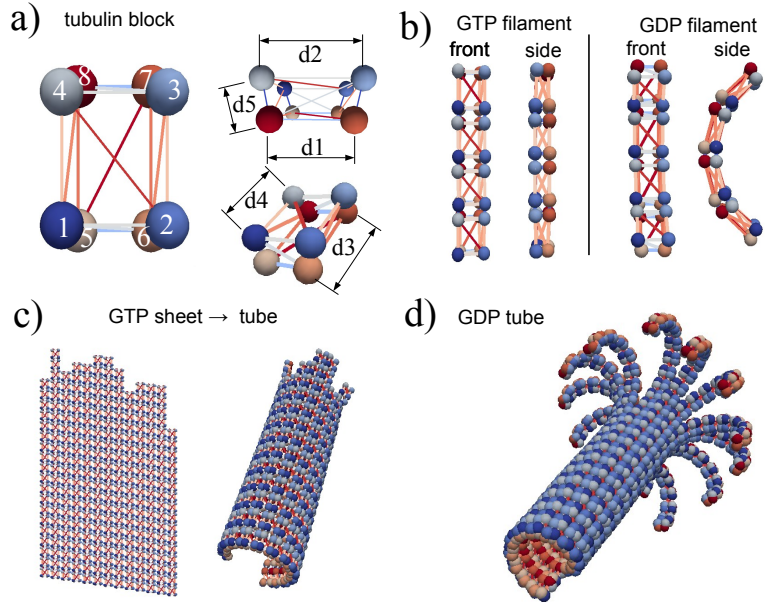} 
\caption{{Three-dimensional tubulin block model of the MT. a) A tubulin block consists of eight nodes connected to neighbouring nodes via stiff linear springs. Diagonal struts are added to avoid shearing and twisting. Each node is also endowed with a hard-core repelling potential. The top and bottom faces of the blocks are of trapezoidal shape to induce lateral curvature
in the MT. b) A PF is obtained by arranging tubulin blocks along a line and connecting them with springs. PFs can be straight or bent, depending on the ratio of top/bottom lengths of the block. c) A sheet of tubulin blocks will form a tubular structure when the opposing sides of the block are trapezoidal. In this case we have $d_2>d_1$ and $d_3=d_4$, corresponding to a GTP bound MT. The sheet will fold with an helicity depending on the number of transversal units and the ratio $d_2/d_1$. d) Clamping one end of the MT while hydrolysing the blocks by letting $d_3>d_4$ and allowing transversal bonds to break leads to the ram's horn shape typical of depolymerizing MTs.}
}\label{fig:block_model}
\end{figure}

We construct a three-dimensional model of a MT starting from a set of wedge-shaped building blocks representing tubulin dimers, as originally suggested in Ref. \cite{hunyadi2007b}. A model similar in spirit to ours has been used in Refs. \cite{cheng2012,cheng2014}
to simulate MT self-assembly. In our model, each block is composed by eight nodes connected by stiff elastic springs, ensuring that the block behaves as a rigid object, as illustrated in \fref{fig:block_model}a. Diagonal springs prevent shearing and twisting of the block, while springs between nearest neighbouring nodes heavily penalize elongation and compression. The rest-length and stiffness of the springs are given in \tref{tab:3dconstants}. By suitably tuning the geometrical features of the tubulin block, we can induce the desired conformation of the PFs and of the MT.

PFs are formed by arranging the blocks linearly --- so that nodes $\#$1,2,5,6 face nodes $\#$3,4,7,8 --- and connecting the nodes by elastic springs with rest length $d_{10}$ and stiffness $k_{10}$, as shown in \fref{fig:block_model}b. When the rest length of the springs on the top face (i.e $d_4$) is equal to the rest length of the springs on the bottom face ($d_3$) the PF is straight (top of \fref{fig:block_model}b), simulating the GTP-bound state. By changing $d_3$ so that $d_4<d_3$ we can induce an intrinsic curvature in the PF, simulating hydrolysis into the GDP-bound state (as shown in \fref{fig:block_model}b).  

The MT is generated by laterally aligning $N$ PFs and connecting the blocks via springs, with rest length $d_{11}$ and stiffness $k_{11}$, to form a sheet as shown in \fref{fig:block_model}c). A tubular structure is obtained by imposing that the top and bottom faces of the blocks have a trapezoidal shape (as shown \fref{fig:block_model}a). In this condition, the PF sheet naturally folds into a cylinder, as shown in \fref{fig:block_model}c. 
We tune the geometrical parameters of the wedge so that a tubulin sheet consisting of $N\approx 11-15$ PFs will close into itself (c.f. \tref{tab:3dconstants}a). A smaller number of PFs will not yield MT closure, while a larger number of PFs produces an overlapping structure. The helicity of the MT lattice is achieved by shifting the alignment of the blocks by $S$ blocks at every turn.
While the most common case is $N=13$ and $S=3$ 	\cite{hunyadi2007b}, here  we consider $N=13$ and $S=2$. 
When the MT closes up we clamp one end of the MT and hydrolyse or add tubulin units at the other end.

The flexural rigidity $B$ of individual PFs in the model can be estimated from the theory of elasticity as the product of the area moment of inertia $I$ and the elastic modulus $E$: $B=EI$. The area moment of inertia of a rod with trapezoidal cross section of height { $h\approx d_5 = 2.5$nm and top and base lengths $d_2=4.55$nm and $d_1=3.5$nm (values correspond to those in Ref. \cite{mickey-howard} and Ref. \cite{vanburen2005} when we consider that the total width and length of a tubulin unit is enhanced by the connecting springs, $d_{10}$ and $d_{11}$, respectively) is $I=d_5((d_2+d_1)/2)^3/12\approx 20$ nm$^4$, } for bending around the median line. 
The elastic modulus is directly related to the stiffness of the springs by $EA= 4 k_{10} d_{10}$, where the factor 4 comes from the fact that we split the tension between two blocks into four parallel springs, and $A$ is the cross section of the block. We chose $k_{10}$ so that $B$ is in the range of {$1.5 - 50 \times 10^{-26}$Nm$^{2}$} which is in agreement with earlier estimates based on measurements of the bending stiffness of MTs \cite{vanburen2005}. Other estimates of the linear angular-spring stiffness of PFs yields slightly different values \cite{molodtsov2}.

Finally, the particles composing the block are endowed with a hard-core repulsion potential (for numerical purposes the repulsive part of a Lennard-Jones potential) with cutoff $r_{hc}=5$nm to avoid inter-penetration of blocks. Longitudinal and transversal bond-lengths ($d_{10}$ and $d_{11}$, respectively) are chosen smaller than the hard-core cutoff to avoid bond-deformation other than stretching. By allowing for bonds breaking and attachment, we can simulate polymerization and depolymerization processes. The model is implemented and simulated using the LAMMPS software package \cite{lammps}.

\subsection*{Kinetochore - Microtubule Interface}

We model the kinetochore-MT interface as an assembly of fibrils \cite{mcintosh-fibrils, mcintosh-MTang}  which represent various canidate proteins and complexes { (e.g. Ndc80, CENP-E, CENP-T, CENP-F.) linked e.g. by the Mis12 complex \cite{dong-ktmt}, the Ska complex \cite{albaabad2014} or the $Dam1$ complex \cite{tanaka-yeast,volkov}. Here the different proteins at the kinetochore-microtubule interface are treated on equal footing -- `coarse grained' into fibrils -- and are modelled as bead-spring polymers}
of length $l_0=20-50$nm \cite{mcintosh-fibrils, mcintosh-ktmt}. We consider hard-core repelling beads of radius $r_b=2$nm connected by linear springs of rest-length $2r_b$ and stiffness $k_{f}=10^6$pN/nm, which makes them practically inextensible.  We also assume that fibrils have vanishing bending stiffness compared to the MTs. Kinetochore fibril bases are arranged randomly on a wide area with a density of $\rho_f=0.77-3.8 \times 10^{-3}$ fibrils/nm$^2$ and then extended towards the incident MT-tip. In practice these densities correspond to 10-50 fibrils per MT. The density was chosen based on the estimate of twenty Ndc80 complexes per captured MT \cite{lawrimore2011}. A summary of fibril parameters is reported in table \ref{tab:3dconstants}b.

\begin{figure}[ht]
\centering
\includegraphics[width=0.7\textwidth]{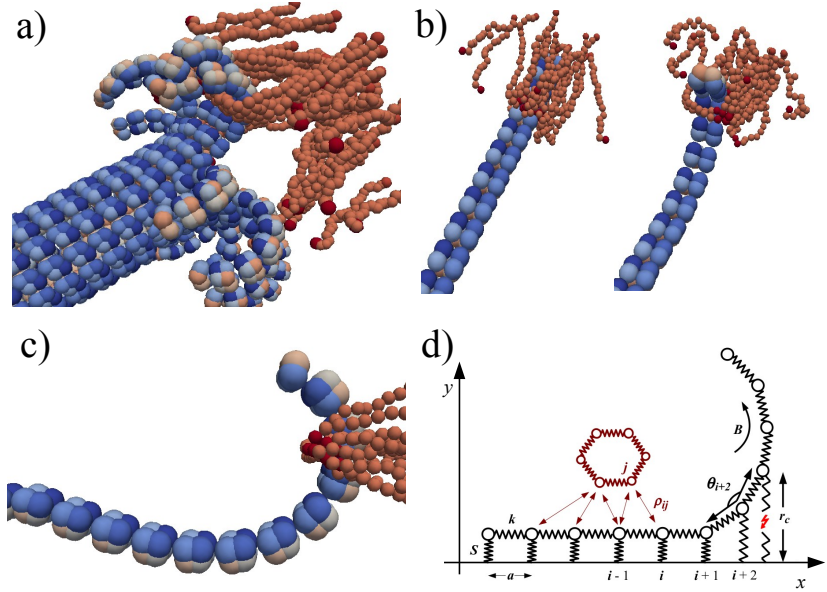} 
\caption{Summary of the model for the kMT interface. a) The fully hydrolysed PFs of the MT curl, forming ram's horns, locking into loops formed by the kinetochore fibrils. b) Genesis of a loop by fibrils attaching to the -- not yet hydrolysed -- tubulin block and locking into each other. When the tubulin blocks are hydrolysed, the curling links the PF tip into the fibrils loop. c-d) Side view of a PF tip with kinetochore-fibril loops around it and its two-dimensional reduction. }\label{fig:kmt_interface}
\end{figure}

Fibril tips can either bind to tubulin blocks in a straight conformation or to other fibril tips. 
In our model, curling tubulin blocks can not directly bind to fibrils, in agreement with experiments showing that the Ndc80 tip binds preferentially with straight tubulin conformations \cite{alushin2010}. This choice is made to
illustrate the worst case scenario, but it is expected that a weak attraction with curved tubulin filaments, possibly relevant for other protein complexes (e.g. CENP-E, CENP-T, Ska, CENP-F), would not change the results.
Hence, an incident MT can attach to kinetochore fibrils either directly by binding to straight tubulin units or  by locking curling PFs into fibril loops as shown in \fref{fig:kmt_interface}b).
In all cases, binding is implemented by a Lennard-Jones potential
\begin{equation}
V_{\rm LJ}(r) = 4\epsilon\left[ \left(\dfrac{\sigma}{r}\right)^{12}-\left(\dfrac{\sigma}{r}\right)^{6}\right] \quad r<r^*
\label{eq:lj}
\end{equation}
where $r^*$ is a cutoff, $\epsilon$ is the binding strength and $\sigma$ controls the width of the potential well.
Notice that fibrils in our model represent coarse-grained protein complexes so that 
their mutual interactions may reflect collective properties. 
 Interaction with GDP-bound tubulin is modelled as a repulsive potential by setting $r^*=\sigma$. {The parameters chosen for $\sigma$ and $r^*$ are chosen so that inter-penetration between fibrils and blocks are avoided} and are reported in Table \ref{tab:3dconstants}c. Movie S1 illustrates the mechanism described here. It shows the result of a simulation where a straight MT enters a bundle of kinetochore filaments and binds some on its outer surface. When the MT protofilaments begin to curl and peel off the MT, some become entangled in loops formed by the kinetochore fibrils.

\begin{table*}[hb]
\centering
\caption{Parameters employed in three dimensional simulations. }\label{tab:3dconstants}
\subcaption{Rest length and stiffness of intra/inter tubulin block springs}
\begin{tabular}{|r|c|c|c|l|}
\hline
Edge & Symbol & Rest length [nm] & Stiffness [pN/nm] & Notes\\
\hline
intra-block & $d_1$ & 3.5 & $k=5\times 10^6$ & incompressible blocks \\
&$d_2$ & 3.5-4.5 & $k$&   \\
&$d_3$ & 5.25-6.3 & $k$& \\
&$d_4$ & 5.25 & $k$&  \\
&$d_5$ & 2.5 & $k$& \\
\hline
inter-block &$d_{10}$ & 1.75 & $k_{10}=(1.5 - 50) \times 10^5$ & calculation based \\
&$d_{11}$ & 1.75 & $k_{11}=50-1000$ & on Refs. \cite{mickey-howard,vanburen2005} \\
\hline
\end{tabular}
\subcaption{Constants used to describe kinetochore fibrils as bead-spring polymers.}
\begin{tabular}{|r|c|c|l|}
\hline
Constant & Symbol & Value & Notes \\
\hline
stiffness & $k_f$ & 10$^6$pN/nm & Taken to be inextensible. \\
bead size & $r_b$ & 2nm & Thin compared to PFs. \\
maximal fibril length & $l_0$ & 20-50nm & Estimated from \cite{mcintosh-fibrils,mcintosh-ktmt}.\\
density & $\rho_f$ & 0.77-3.8 $\times 10^{-3}$ fibrils/nm$^2$ & Estimate, Ref. \cite{lawrimore2011}.\\
\hline
\end{tabular}
\subcaption{Lennard-Jones parameters for interactions of fibril tips.}
\begin{tabular}{|r|c|c|c|l|}
\hline
Interaction &  $\epsilon$ [pN nm] & $\sigma$ [nm] & $r^*$ [nm] & Notes \\
\hline
Fibril-tip - fibril-tip  & 100-2000 & 2.5 & 3.75 & Assumption: strong attraction.\\
Fibril-tip - straight PF  & 1.5-5.00 & 7.5 & 30 & Assumption: weak attraction.\\
Fibril-tip -  curved PF  & 500 & 7.5 & 7.5 & Assumption: repulsion.\\
\hline
\end{tabular}

\end{table*}

\subsection*{Simplified two-dimensional model of MT-protofilament elasticity}\label{sec:2dmodel}

The three-dimensional model of the MT and the MT-kinetochore interface is highly complex and difficult to simulate for long times due to the stiffness of the bonds. Therefore only limited results can be obtained by the full 3D model. To obtain relevant statistics 
for the detachment kinetics, we need to perform some important approximation reducing the problem to two-dimensions. \Fref{fig:kmt_interface}d suggests a way how such a dimensional reduction might take place. The protofilament is modelled as a flexible chain of nodes connected via springs \cite{zapperi-mahadevan} with longitudinal stiffness $k=15000$pN/nm and rest length $a=10$nm (approximately one tubulin dimer).
Since the fibrils form loops around the PF tip, the only way to detach -- save the dissociation of the fibril loop -- is by sliding the loop off the PF ram's horn. In two dimensions, this process is illustrated in \fref{fig:kmt_interface}d) where the cross-section of the fibrils are modelled as an elastic ball.

Here, we model the kinetics of single protofilament attached to a substrate representing the interaction with neighboring protofilaments \cite{janosi2002,zapperi-mahadevan}.  The chain furthermore has an intrinsic bending angle of $\varphi$ \cite{mcintosh-MTang} and bending stiffness $B$. 
 Hence, the elastic energy of a protofilament with node coordinates $\mathbf{r}_i$  is given by
\cite{zapperi-mahadevan}
\begin{eqnarray}
\mathcal{E}_{el,PF} = \sum_i \frac{1}{2}k(|\mathbf{r}_{i+1}-\mathbf{r}_i|-a)^2
-B/a \cos(\theta_i-\varphi)+H(r_c-y_i)\frac{1}{2}Sy_i, \label{eq:elasticity}
\end{eqnarray}
where $H(x)$ is the Heaviside step-function and $\theta_i$ are the angles between neighbouring subunits. Hence, the elastic force on each node $i$ is simply $\mathbf{f}_{el,PF}^i=-\partial \mathcal{E}_{el,PF}/\partial \mathbf{r}_i$. 

As in the 3D model, we estimate the bending stiffness of the PF from the flexural rigidity of a microtubule $EI_{\rm MT} \approx 10 \times 10^{-24}$Nm$^2$ (see \cite{vanburen2005} and references therein) and its second moment of inertia $I_{\rm MT}\approx 2\times 10^{-32}$m$^4$ \cite{mickey-howard}. Although the protofilament is modelled as a one-dimensional object, we estimate its model of inertia treating it as a cylinder of radius  $b=5$nm, which yields  $I_{\rm PF}\approx  10^{-34}$m$^{4}$. By simple calculation we can then estimate $B=(EI_{\rm MT}/I_{\rm MT})I_{\rm PF}\approx 5\times 10 ^{-26}$Nm$^2$. 
The interaction between the filament and the substrate is modelled by a spring with stiffness $S=1000$pN/nm which breaks when it is extended beyond $r_c=1$nm . These parameters ($S$ and $r_c$) were fitted to the requirement that depolymerizing MTs shrink at the experimentally observed speed, keeping a finite curvature at the tip to capture the fibrils. In fact, as shown in Ref. \cite{zapperi-mahadevan}, only the combination $Sr_c^2$ rules the PF dynamics.  

\subsection*{Two dimensional simulations of fibril-MT interactions}


As shown in \Fref{fig:kmt_interface}d we consider only the cross section of the fibril bundle (Fig. \ref{fig:kmt_interface}c) and model it a  closed chain of six nodes connected by linear and angular springs with extension $b$, stretching stiffness $k_f$ and bending stiffness $B_f$. The elastic energy of the chain can then be written in a form equivalent to \eref{eq:elasticity}.

The interaction between the PF and the fibril is modelled by a repulsive potential in case of a depolymerizing
MT and by an attractive force for a polymerizing MT:
The repulsive force between for each node of the depolymerizing PF $\mathbf{r}_i$ and the fibril $\mathbf{R}_j$  according to
\begin{eqnarray}\label{eqn:interaction}
\mathbf{f}_{int}^{ij}= - A \frac{\exp(-0.1\rho_{ij})}{\rho_{ij}^3} \mathbf{r}_{ij} \nonumber
\end{eqnarray}
where $\mathbf{r}_{ij}=\mathbf{r}_i-\mathbf{R}_j$ and  $\rho_{ij}=|\mathbf{r}_{ij}|$. 
The attractive interaction with polymerizing PF is modeled according to Eq. \ref{eq:lj}.

\subsection*{Numerical simulations of fibril-protofilament dynamics}

The dynamics of the PF-fibril system is governed by the following coupled Langevin equations 
\begin{subequations}
\begin{align}
\gamma_{PF} \frac{d \mathbf{r}_i}{dt}&= \mathbf{f}_{el,PF}^{i}+\mathbf{f}_{int}^{ij} \label{eq:ODa}\\ 
\gamma_{f} \frac{d \mathbf{R}_j}{dt}&= \mathbf{f}_{el,f}^{j}-\mathbf{f}_{int}^{ij} + \mathbf{F}+\mathbf{G}_j\label{eq:ODb}, 
\end{align} 
\end{subequations}
where $\gamma_{PF}$ and $\gamma_{f}$ are the damping coefficients of the nodes of the PF and the fibril, respectively,
and $\mathbf{F}= F \hat{x}$ is an externally applied pulling force, acting along the horizontal direction. The thermal  noise $\mathbf{G}_j$ acting on the fibril is a Gaussian random  force with  average $\langle \mathbf{G}_j(t)\rangle=0$ and correlations $\langle \mathbf{G}_j(t)\mathbf{G}_k(t')\rangle=2\gamma_f k_{\rm B}T\delta_{ij}\delta(t-t')$, where $T$ is the temperature and $k_B$ is Boltzmann constant. In the following we quantify the amplitude of thermal fluctuations by the parameter $\omega\equiv\sqrt{2 k_{\rm B} T/\gamma_f}$. 

The damping coefficient can be evaluated considering that the drag coefficient of a sphere of radius $r$ immersed in a fluid with dynamical viscosity $\eta$ is given by the Stokes formula $\gamma=6\pi\eta r$. The dynamical viscosity of water at  $T \approx$ 300K is $\eta\approx 10^{-3}$ Pa s. Assuming that  the radii of the fibril and PF are $r_f=1$nm and $r_{PF}=10$nm, we get $\gamma_f\approx 10^{-10}$ kg/s and $\gamma_{PF}\approx 10^{-9}$ kg/s. In these conditions, the thermal fluctuations of the PF are overshadowed by the interactions with the strongly fluctuating fibrils and can be safely dropped. This is particularly convenient because it further speeds up the simulations. 

The MT shrinks or grows  by adding or removing  tubulin subunits from the tip, but in mammalian cells,  MTs also depolymerize at the spindle poles which is associated with poleward flux of tubulin \cite{mitchison-flux}.
Here, we consider that the PF is able to polymerize (attach subunits at the tip) or depolymerize (lose subunits from the tip) and switch between these states. The growing/shrinking velocities and switching rates we use are taken from experiments in refs \cite{akiyoshi, rusan, tirnauer2002}. 
We  assume that the velocities are dependent on the force on 
the tip of the protofilament \cite{akiyoshi} according to
\begin{equation}\label{eqn:arrest}
v_{\rm g/s}(F) = v^0_{\rm g/s}\times 10^{\pm F/F_{\rm g/s}}.
\end{equation}

Eqs. \ref{eq:ODa} and \ref{eq:ODb} are solved by a fourth-order Runge-Kutta algorithm for the PF 
and an Euler-Maruyama algorithm for the fibril. Due to numerical constraints, it is not possible to
simulate the model over a timescale that is comparable to the experimental one. We thus 
artificially increase the noise fluctuations and then analyze how the results depend on $\omega$.
In this way, we are able to show that the numerical results approach the experimental ones
as $\omega$ tends towards realistic values. The numerical values of the constants used in the simulations 
are summarized in table \ref{tab:constants}

\setcounter{table}{1}
\begin{table*}[ht]
\centering
\caption{List of constants used in two-dimensional simulations.}\label{tab:constants}
\begin{tabular}{|p{4.3cm}|c|c|p{3cm}|}
\hline
Name & Symbol &Values used &Comment \\
\hline
\multicolumn{4}{|l|}{\bf MT-Protofilament} \\
\hline
subunit length & $a$ & $10^{-8}$ m & length scale \\
drag coefficient & $\gamma_{PF}$ & $10^{-9}$ kg/s & estimate \\
bending stiffness & $B$ & $5 \times 10^{-26}$ Nm$^2$ & calculation  \\
stretching stiffness & $k$ & $1.5 \times 10^{-24}$ N/m & assumption \\
substrate stiffness &$S$ & $1.0 \times 10^{-26}$ N/m & fit \\
breaking length &$r_c$ & 1nm& fit \\ 
bending angle & $\varphi$ & 0.18 -0.42 & \cite{mcintosh-MTang} \\
growth velocity & $v^0_{\rm g}$ &0.008- 0.21 $\mu$m/s & \cite{rusan,akiyoshi,tirnauer2002}\\
shrinking velocity & $v^0_{\rm s}$& 0.2-0.3 $\mu$m/s & \cite{rusan,akiyoshi} \\
growth acceleration & $F_{\rm g}$ & 20pN & \cite{akiyoshi} \\
shrinking stall & $F_{\rm s}$ & 7pN & \cite{akiyoshi} \\
rescue rate & $1/p_{\rm res}$& 0.045/s & \cite{rusan}\\
catastrophe rate & $1/p_{\rm cat}$ &0.058/s & \cite{rusan}\\
\hline
\multicolumn{4}{|l|}{\bf Kinetochore fibril}\\
\hline
segment length & $b$ & $10^{-9}$ m & assumption \\
drag coefficient & $\gamma_f$ & $10^{-10}$ kg/s & estimate \\
flexural rigidity & $B_f$ & $5 \times 10^{-27}$ Nm$^2$ & assumption  \\
stretching stiffness & $k$ & $10^{-25}$ N/m & assumption \\
fluctuation amplitude & $\omega$ & 15-35 nm $\mu$s$^{-1/2}$ & tried values\\
\hline
\end{tabular}
\end{table*}

\section*{Results}

\subsection*{Microtubule protofilaments entangle with kinetochore fibrils}

\begin{figure*}
\centering
\includegraphics[width=0.5\textwidth]{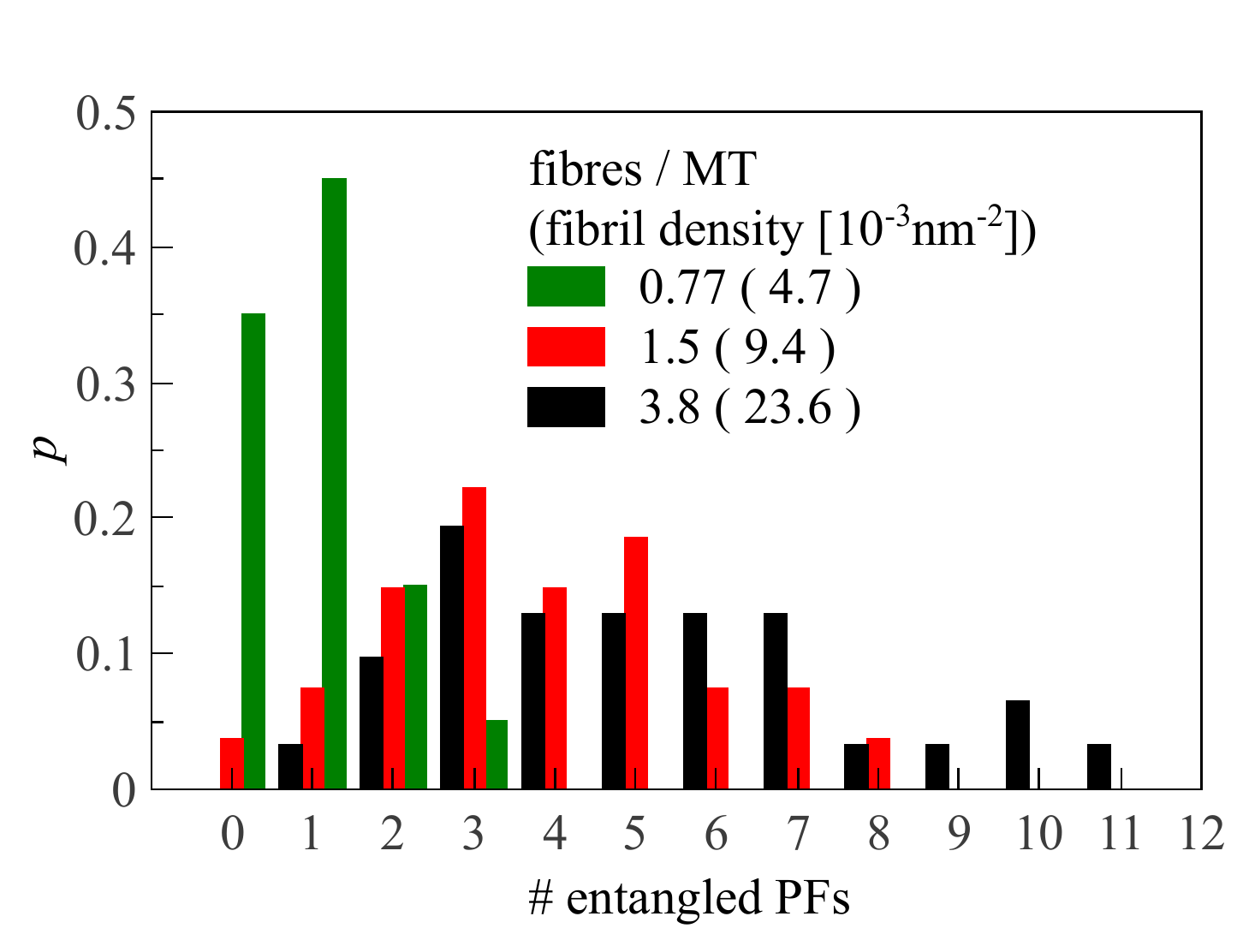}
\caption{{Histogram of the number of entangled PFs interacting with a set of kinetochore fibrils for three different
densities. The total number of simulations is 30 in all cases. For details see text.}}\label{fig:histogram}
\end{figure*}

We simulate the three dimensional kinetochore-MT interface model presented and discussed in the methods section. In \fref{fig:kmt_interface}a) we show a MT whose curling PFs are partly linked into loops formed by kinetochore fibrils. In \fref{fig:kmt_interface}b) and Movie S2 we illustrate the process by which a curved PF locks into a loop formed by kinetochore fibrils.  A polymerizing PF, composed by straight tubulin blocks, reaches into the entangled kinetochore outer plate and fibrils attach to the tubulin blocks (see \fref{fig:kmt_interface}b) left). Since fibril tips are now close together, they can easily bind with each other (as suggested in Ref. \cite[Figure 6C]{ciferri2008} and Ref. \cite{zaytsev2013}). When the PF depolymerizes and starts to curl, fibrils detach from the tubulin due to the reduced affinity, but the ram's horn of the PF can lock mechanically into the loop formed by the connected fibrils (see \fref{fig:kmt_interface}b) right).  

To test the fidelity of out kinetochore-microtubule interface model, we perform a series of simulations to see how many PFs are entangled in the fibril loops.   The MT-tip was kept straight and moved into the mass of kinetochore fibrils with a velocity of 1$\mu$m/min \cite{tirnauer2002}. Fibrils attach to the MT surface and bind to each other, especially when they are in proximity of the MT, thus forming loops. As soon as the tip of the longest PF reaches the area of the fibril bases, the incident velocity is set to zero, and the MT is set to depolymerize immediately. In this process curling PFs can hook into the fibril loops. 
We perform this simulation for various fibril densities and average over 30 realizations in each case. We report the number of entangled PFs as a histogram in Fig. \ref{fig:histogram} for 10, 20 and 50 fibrils per MT, or 0.77, 1.5 and 3.8 fibrils per PF. 
The average number of entangled PFs is 1.1, 4.5 and 5.2 for 10,20 and 50 fibrils, respectively. Hence even for a relatively small
number of fibrils, there is significant and robust entanglement.

The geometrical conformations of the PF tips broadly correspond to the two distinct states of the dynamic MT, a growing/polymerizing (straight) and a shrinking/depolymerizing (curved) state, with differing chemistry and physics of the molecules at the tip. In the growing state, the tip of the MT recruits phosphorylysed GTP-tubulin subunits, while the tubulin subunits in the rest of the MT body are progressively hydrolysed. The main difference -- for our purposes -- between GTP and GDP-tubulin is the equilibrium angle between two connected subunits. While a filament made out of GTP units tends to remain straight, a GDP-tubulin filament has a curved equilibrium configuration (see Fig. \ref{fig:block_model}b). As a consequence, the tips of PFs in a growing MT will tend to have rather straight configurations while in the depolymerizing phase, the PFs of the MT form hooks - ram's horns - while losing tubulin subunits. 

To estimate the load carrying capacity of the MT-kinetochore interface, we consider only  the tip of the PFs and assume that the constituting tubulin blocks are uniformly either hydrolysed (GDP-bound) or phosphorylated (GTP-bound). In this way, we can discuss detachment from depolymerizing or polymerizing PFs separately and distinguish the underlying detachment mechanisms.  In particular, a single PF is first inserted, up to a few blocks, typically ranging from one to three, into a network of kinetochore fibrils and the system is then relaxed for 10s. Since we always start from a GTP-tubulin PF, the tips of some fibrils bind to the ``outer" side of the tubulin blocks (particles with $\#$1-4, c.f. \fref{fig:block_model}a). When the system is equilibrated, the distal end of the PF is held fixed, while the  kinetochore plate, to which fibrils are attached, is moved with constant velocity $v=10$nm/s. In these conditions, we measure the longitudinal component (parallel to the pulling direction) of the interaction force between kinetochore and the PF.

\begin{figure}[H]
\centering
\includegraphics[width=0.8\textwidth]{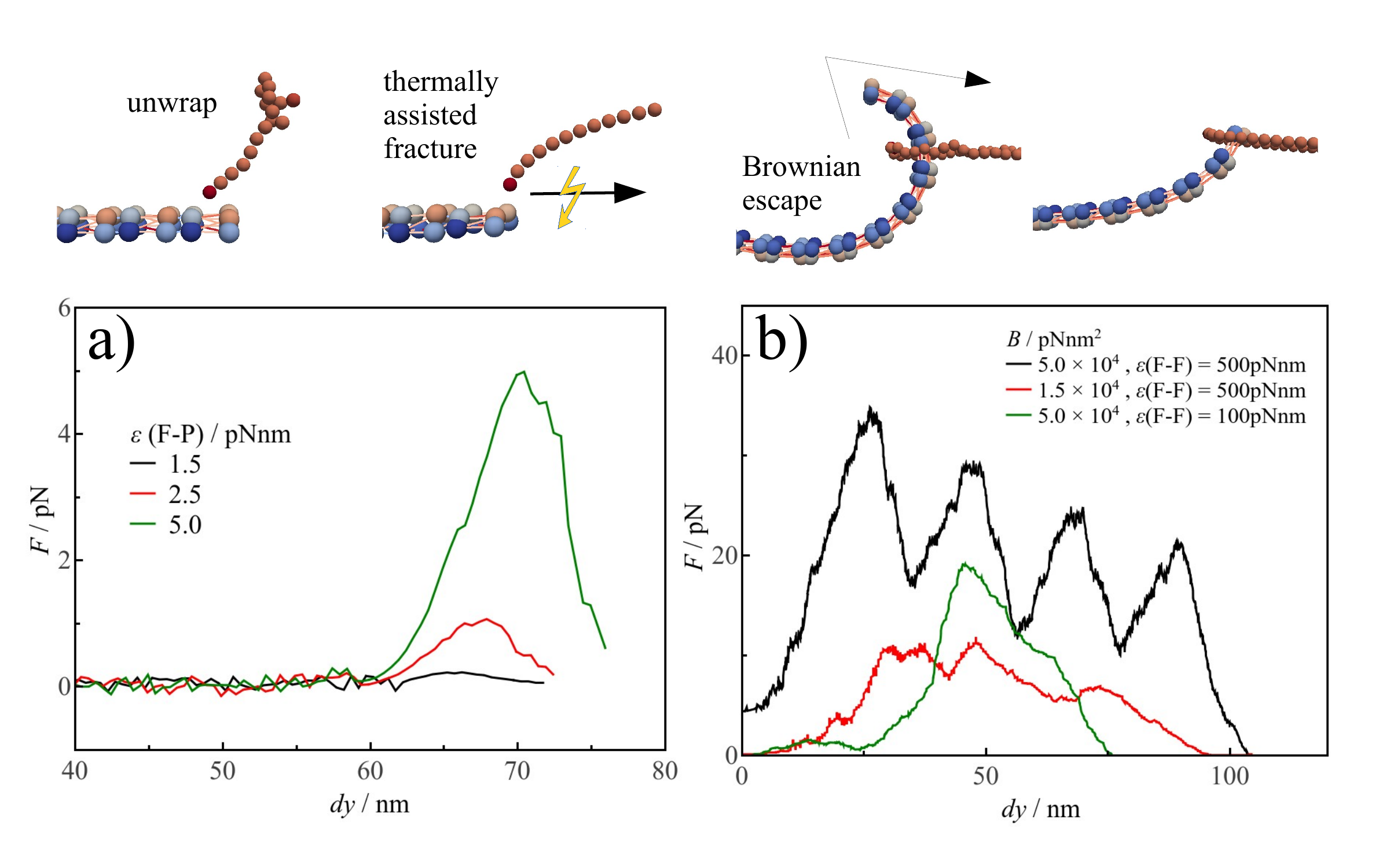}
\caption{Detachment mechanisms of the kinetochore-PF interface. a) Force-displacement curve of a straight PF-kinetochore fibrils interface. The fibrils are attached to the first tubulin block of the PF because the fibril tips interact with the PF via a Lennard-Jones potential with strength $\epsilon$(F-P). Displacing the fibril ends leads at first to no increase of the force as long as the fibril unwraps. Upon completion of this straightening process, the force increases rapidly as the fibrils and the PF stretch while overcoming Lennard-Jones attraction. b) Force-displacement curve of a GDP-tubulin PF in the curved conformation. Again, the kinetochore fibrils are pulled with constant velocity and the in-line component of the interaction force is plotted for flexural rigidity $B=5\times 10^4$ pNnm$^2$  and $B=1.5\times 10^4$ pNnm$^2$. The interaction energy between the fibrils is modelled
by a  Lennard-Jones potential with $\epsilon$(F-F)$=500$pNnm or $\epsilon$(F-F)$=100$pNnm. In the first case the peak corresponds to the bending of the PF while in the second case to the breakdown of the fibril loop.
The Oscillation of the black curve is due to the discrete nature of the PF and arises from the sliding of the kinetochore fibrils from one block to the next. This oscillation is drowned out by noise when the flexural rigidity is reduced as shown in the red curve.}
\label{fig:x}
\end{figure}

For GTP-tubulin PFs, we plot the force-displacement curves in \fref{fig:x}a for three values of the binding strength  $\epsilon$(P-F) between fibrils and PFs. The fibril tips are in this case attached only to the first tubulin block of the PF in a way shown in the inset. The curves show the displacement $dy$, the difference between the starting and end positions of the kinetochore plate as a function of the in-line component of the interaction force, normalized to the number of attached fibrils. Simulations are repeated 30 times to average out the effect of thermal noise. The results show that at short displacements the kinetochore plate feels no restoring force since the fibrils are ``unwrapped", i.e. straightened. At this point, attached fibrils finally exert a force on the plate due to the binding with the tubulin dimer, until the bond breaks. The detachment from a GTP-tubulin PF, therefore relies \emph{solely} on the attractive interaction between kinetochore fibrils and tubulin blocks. The process is illustrated in Movie S3.

When the PF is formed of GDP-tubulin, its end is curved and the fibril-tubulin interaction is turned into a purely repulsive potential. We perform the same simulation as detailed for the GTP-tubulin PF above. The difference is that the PF is inserted more than one tubulin-block into the fibril network, typically two to four blocks, and the relaxation also entails the curling of the PF. We performed simulations for two values of the flexural rigidity and plot the results in \fref{fig:x}b (see also Movie S4). For the larger value of the stiffness we observe pronounced oscillations of the force-displacement curve due to the sliding of the kinetochore fibrils on the tubulin blocks while uncurling. For the less stiff case, the oscillations are not clearly discernible and the 
detachment of the kinetochore-MT interface is much smoother.  As shown in Fig. \ref{fig:x}b, the peak load of a single depolymerizing PF is larger than 10 pN, suggesting that a MT can on average carry a peak load of up to 50pN, which compares nicely with experimental results \cite{grishchuk-MTs}. We also test the effect of the binding interactions between fibrils $\epsilon$(F-F). For $\epsilon$(F-F)=100 pNnm and $B=5\times 10^4$ pNnm$^2$, the fibril loop eventually breaks (see Movie S5) but it is still able to carry a load of $20pN$.
Hence for $\epsilon$(F-F)=100 pNnm and $B=1.5 \times 10^4$ pNnm$^2$, the fibril loop should be strong enough to sustain a load of 20 pN, as we verified numerically.


A feature of the detachment that is not captured in the MD simulations is thermally induced/assisted detachment prevalent at low forces. The reason is that thermal phenomena in a system as ours appear on time scales much larger than we are able to simulate
in three dimensions. To be able to simulate on statistically relevant time scales we developed  and implemented a two-dimensional equivalent of the kinetochore - MT interface as already discussed in a previous section.

\subsection*{Shrinking microtubules form catch-bonds with kinetochore fibrils due to their conformation}

For a shrinking PF, we assume that the fibrils do not attract to the tubulin molecules that are in curved conformation \cite{alushin2010} but interact via a solely repulsive potential, as discussed in the methods section.
To relate the experiments of Ref. \cite{akiyoshi} with the two-dimensional model, we chose an intrinsic bending angle of $\varphi=0.36$,  corresponding approximately to the peak of the curvature distribution measured experimentally in \cite[figure 4c]{mcintosh-MTang}.  All the model parameters are reported in Table \ref{tab:constants}.

\begin{figure}[H]
\centering
\includegraphics[width=0.97\textwidth]{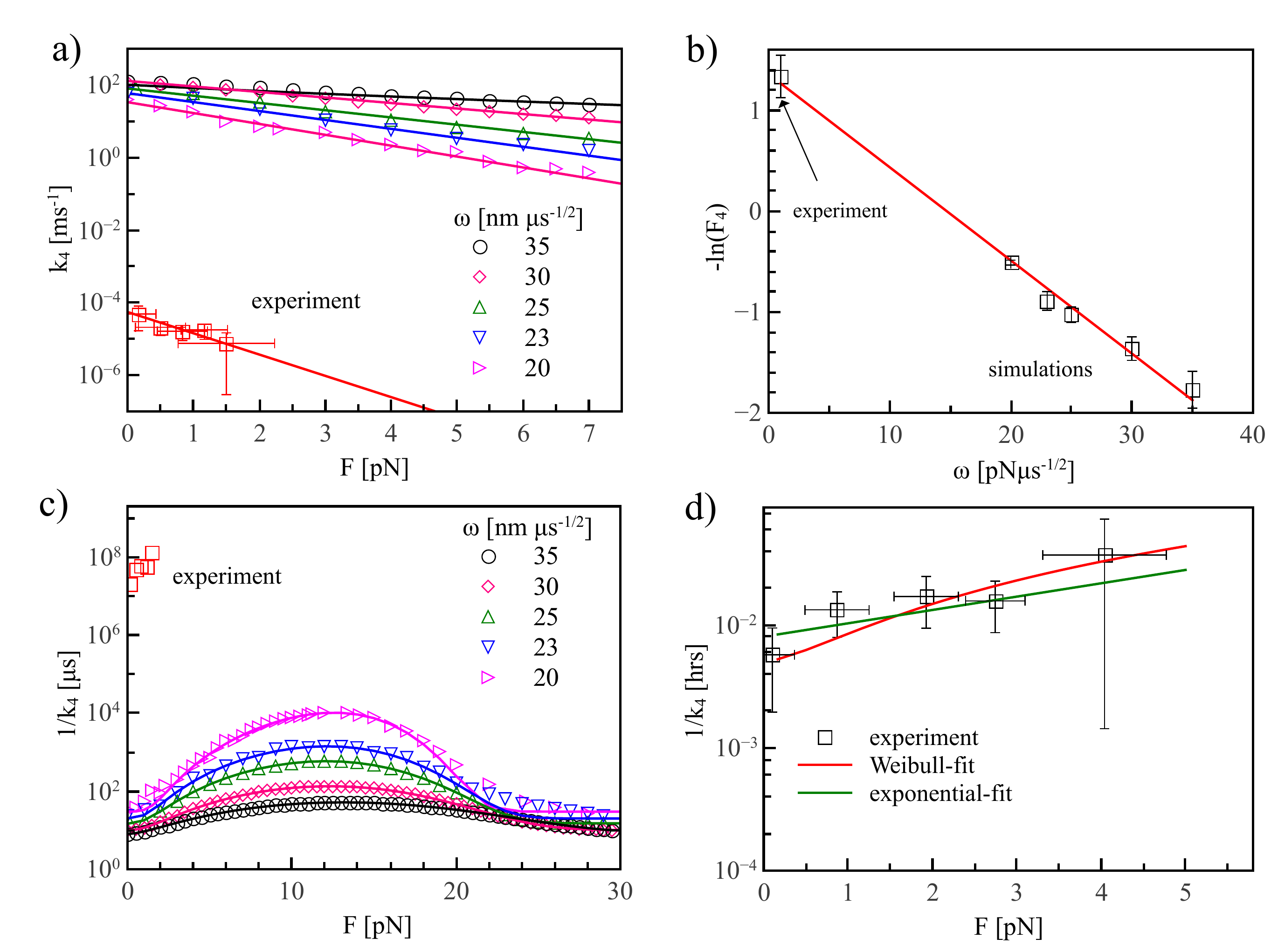}
\caption{{Simulation of the catch-bond behavior for depolymerizing PFs.} a) The detachment rate as a function of the 
pulling force for different values of the diffusion parameter $\omega$ for the 2D model of a depolymerizing PF together with the experimental data from Ref. \cite{akiyoshi}. b) Extrapolation of the diffusion parameter $\omega$ in the experiments from the numerical data.
c) Attachment lifetimes (inverse detachment rates) of depolymerising PFs to kinetochores at high forces. The numerical data suggests that the lifetime peaks due to deformations and then decreases sharply with applied load. The low-force parts and the peaks can be fitted well with the Weibull distribution which yields also a reasonable fit for the experimental data, as shown in panel d).
}\label{fig:2}
\end{figure}

As described in the previous section, detachments can occur either because of thermal and other non-equilibrium fluctuations that lead the fibril to jump off the PF tip, or because a sufficiently strong pulling force mechanically bends the PF. In agreement with Ref. \cite{akiyoshi}, low  pulling forces increasingly stabilize the PF-fibril attachment. Furthermore, our model predicts that for larger forces the attachment should again be destabilized. To compare the model with experiments, we plot in Fig. \ref{fig:2}a the numerical and experimental data for the detachment rate, denoted  $k_4$ as in Ref.\cite{akiyoshi}. The numerical data is plotted for various values of the fluctuation parameter $\omega$. At small forces the detachment rates is well fitted by exponential decay, $k_4=k_4^0\exp(-F/F_4)$. The parameter $F_4$ follows an exponential curve as a function of $\omega$, as shown in Fig. \ref{fig:2}b. From this plot we can extrapolate the value of $\omega$ in the experiment for our model and arrive at $\omega\approx 1 \rm{nm}\mu \rm{s}^{-1/2}$ which is a realistic and physical fluctuation parameter for this system.  Hence, at low forces simulations and experiments fit neatly together.

It is, however, unlikely that bond strengthening persists up to larger forces, where elastic deformation of the PF should take place.	Indeed, simulations for large forces show that the lifetime of the kinetochore-microtubule attachment -- the inverse of $k_4$ -- displays a peak at $\sim$ 10 pN, and decreases very quickly after that, as shown in Fig. \ref{fig:2}c. The detachment times can be very well fitted at low forces and around the peak with a Weibull function
\begin{equation}
1/k_4(F)=\tau^0+A_0\frac{k}{\lambda}\left(\frac{F}{\lambda}\right)^{k-1}{e}^{-\left({F}/{\lambda}\right)^{k}}, 
\end{equation}
where $\tau^0$ the lifetime at zero force, the amplitude $A_0$, and the Weibull parameters $\lambda$ and $k$ all depend on
a diffusion parameter $\omega$. The Weibull function provides also an alternate fit for the lifetime of attachments with depolymerising PFs in the experimental data as shown in Fig. \ref{fig:2}d, together with the original exponential fit. Both fits yield consistent results but it is difficult to decide which one is better.
In both cases, the reduced $\chi^2$ resulting from least-square minimization is of the order of $10^{-5}$
which is a clear indication of over-fitting.

Our model suggests that  measuring the lifetime of depolymerizing MT attachments to kinetochores should result in a peak at
larger forces. As mentioned before, detachments at large force are due to a different physical process than 
the detachments at low forces. In Ref. \cite{akiyoshi} the destabilization of the kinetochore-MT attachment was attributed to a switch of the MT from the shrinking to the growing state. While we do not discount that possibility, we suggest the existence of an alternate pathway which could be dominant at short time scales: When the pulling force is large enough, the depolymerizing PF is uncurled (c.f. movie S2). 

\subsection*{{Growing microtubules display force weakening bonds ruled by the binding affinity with kinetochore fibrils}}
\begin{figure}[H]
\centering
\includegraphics[width=0.8\textwidth]{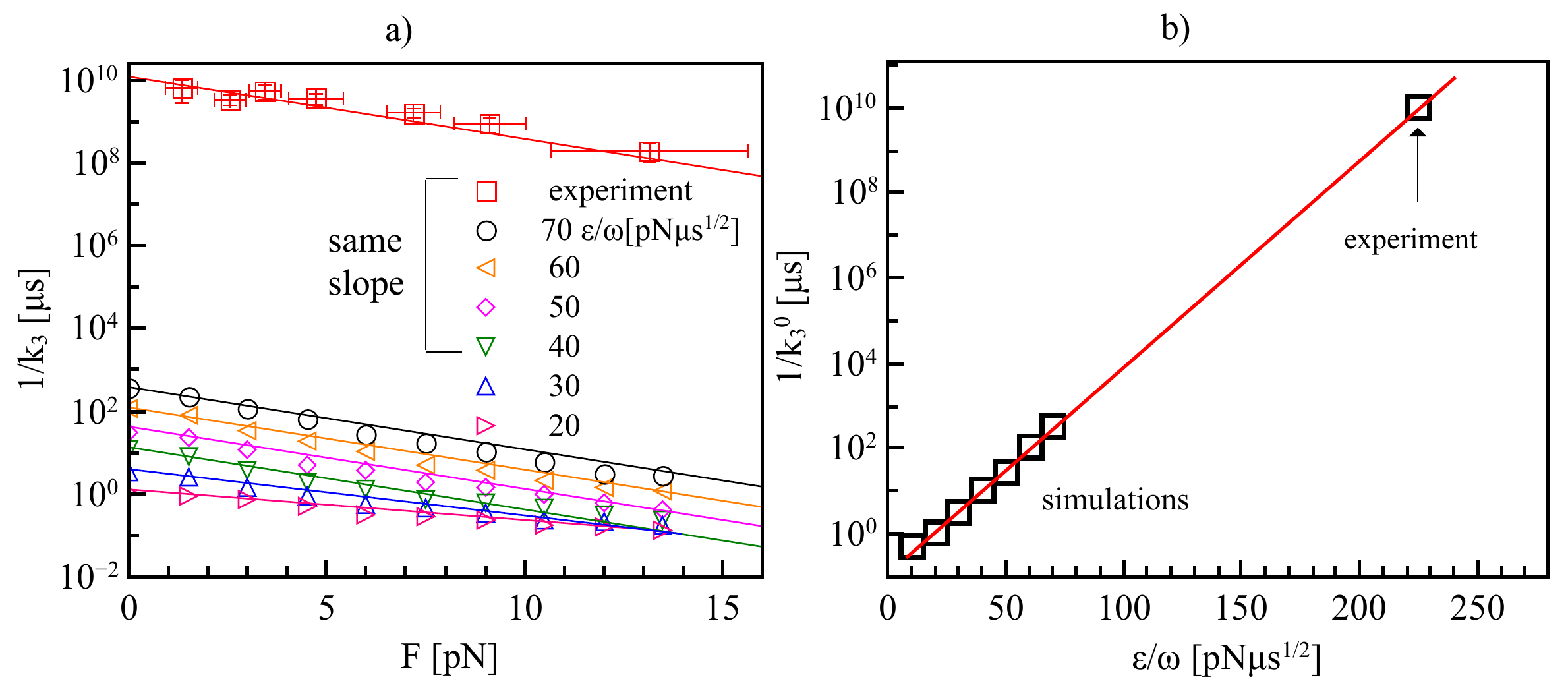}
\caption{Detachment from a growing PF. a) Attachment lifetimes obtained for the 2D model for polymerizing PFs as a function of the binding strength-fluctuation ratio $\epsilon/\omega$ together with the experimental data from Ref. \cite{akiyoshi}. For high enough binding strengh/fluctuation ratio, the parameter $F_3$, appearing as the slope here, stays constant and is the same for the numerical and experimental data. b) Extrapolation of the experimentally observed binding strength. }
\label{fig:k4}
\end{figure}
In the two dimensional model, kinetochore fibrils are pulled with a constant force, while the PF is growing or shrinking. The time-frame for switching between the two states is much larger than the detachment times we are able to simulate and therefore the effects of rescue and catastrophe do not affect our numerical results. To further simplify the numerical simulation and thus results for longer times, we model the detachment from polymerising PFs as a simple escape from a Lennard-Jones potential with a cutoff and simulate Eq. \ref{eq:ODb}. We have tested in some specific cases that the results are indistinguishable from those obtained with
the complete model.

The result for various values  is plotted in 
Fig. \ref{fig:k4}a reports the value of the attachment lifetime --  denoted by $k_3$ as in Ref. \cite{akiyoshi} -- obtaiend
in simulations as a function of the binding energy-fluctuation ratio $\epsilon/\omega$ together with the experimental
data. Numerical simulations and experimental data are well fitted with exponential curves and have the same slope on a logarithmic scale. In particular, we fit the rate as
\begin{equation}
k_3(F)=k_3^0(\epsilon/\omega)\exp(F/F_3),
\end{equation}
where only the detachment rate at zero force $k_3^0$ depends on the ratio $\epsilon/\omega$, while $F_3\approx$ 3.8 pN is the same for experimental and numerical data. Plotting $1/k_3^0$ versus $\epsilon/\omega$ for the numerical result, allows us to extrapolate the value for $\epsilon/\omega$ for the experiment to $\epsilon/\omega\approx 225 \rm{pN}\mu \rm{s}^{1/2}$, as shown in Fig. \ref{fig:k4}b. Since we previously estimated $\omega\approx 1\rm{nm}\mu\rm{s}^{-1/2}$ for the experiment, we predict the attachment strength of the fibrils to the kinetochore to be $\epsilon\approx 225 \rm{pNnm} \approx 50 k_{\rm B} T$. This is of the order of magnitude for usual protein-protein interactions, providing an additional confirmation of the validity of our approach.

\subsection*{Protofilament intrinsic curvature can determine the stability of mi\-cro\-tu\-bu\-le-kinetochore attachments}
The main point of cell division is the segregation of replicated chromosomes into two daughter cells. This is chiefly executed by the mitotic spindle, consisting of MTs, organized into two astral structures whose centres derive from centrosomes and associated proteins. Faithful segregation of the chromosomes requires proper bi-orientation and thus dynamic kinetochore-MT attachments, i.e. it must allow erroneous attachments -- kinetochores attached to a MT emanating from the wrong pole -- to be corrected. In other words, kinetochore-MT attachments can be unstable or stable, short-lived or long-lived.

For a given value of the diffusion amplitude $\omega$ we can vary $\varphi$ and plot the detachment times with respect to constant pulling force, we see that the peak height as well as the offset $\tau_0$ decrease with $\varphi$, and at a critical value the detachment time at zero force $\tau_0=t(F=0)$ is higher than the peak. Decreasing $\varphi$ further makes the peak disappear completely and we end up with a detachment curve that is purely exponentially decreasing with applied force. 
As an example, we plot in Fig. S1a the detachment times for $\omega=20$nm $\mu$s$^{-1/2}$ for set of values of $\varphi$. As $\varphi$ decreases, the catch bond peak is reduced and is shifted to smaller forces. At a critical value - for $\omega=20$nm $\mu$s$^{-1/2}$ it is $\varphi\approx 0.275$ - the peak detachment time is at the same height as the zero-force detachment time $\tau_0=t(F_{\rm peak})$. Then, the peak completely disappears for $\varphi<0.15$ and the detachment curve is a simple exponential decay. These observations allow us to construct a very simple ``phase diagram" where the transition line corresponds to the condition for which
the catch bond disappears.  In the stable region the attachment time increases with small applied forces, while
it decreases in the the unstable region. An example for $B=5\times 10^4$pNnm$^2$ and $\omega=20$ is  shown in Fig S1b.  A stable region for $\varphi>0.275$ (right dash) and an unstable region for $\varphi<0.275$, which splits into a ``weak" part (left dash) where $\tau_{\rm peak}<\tau_0$ and a truly unstable part (cross dash) with no peak at all. The same procedure can be repeated for different values of $\omega$, allowing us to construct a phase diagram for the stability of the attachment which we report in Fig. S2.

\section*{Discussion}

\subsection*{A conformational mechanism for the catch-bond}

Our model for MT-kinetochore attachments captures and explains in easy-to-understand terms several key features of experimentally observed phenomena. Most prominently, we could reproduce the detachment times of MTs from kinetochores that are pulled with constant force \cite{akiyoshi}. Using the model we simulate the emergence of the {\it catch-bond} -- the stabilization by tension -- of the kinetochore interface with a depolymerizing protofilament,
emphasizing the role of the geometrical conformation of the kinetochore-microtubule interface.

We identified different mechanisms of kinetochore-microtubule detachment for straight and curled protofilament tips.
\begin{enumerate}
 \item[i] Kinetochore fibrils are attached to straight protofilaments and detach due to thermal activation. This detachment is facilitated by an applied force, resulting in exponential weakening of the attachment.
 \item[ii] Neighbor fibrils that have attached to each other form loops into which curling protofilaments hook.  When no external force is applied the loop can escape by thermal fluctuations, i.e. Brownian motion.
 \item[iii] When the loop and PF-tip are pulled apart, the thermal fluctuations acquire a bias to move preferentially towards the tip of the PF, making an escape from a ram's horn conformation more unlikely, thereby stabilizing the attachment. However, if the pulling force is too strong, then the ram's horn becomes uncurled or the loop breaks, so that the attachment is destabilized again.
\end{enumerate}
 
The third point makes an experimental prediction -- at higher forces the detachment rate from depolymerizing PFs increases again -- that could be easily be confirmed experimentally. Accordingly, one should extend the experiments plotted in Figure 4c of Ref. \cite{akiyoshi} to measure the detachment rate from depolymerizing MTs at higher forces. The two-state model of Akiyoshi et. al. implicitly assume that the detachment rate should continue to decrease with applied force, while our model predicts that it should eventually go up.

\subsection*{Geometrical aspects of MT-kinetochore attachment stability}

Another prediction that follows from our results is the stabilization of kinetochore-MT attachments via the intrinsic curvature of the MT protofilaments. The simulations show that the higher the PF curvature the longer the lifetime of the attachment. If the intrinsic curvature is  small enough then the stabilization, together with the {\it catch-bond} disappears and the attachment becomes weaker with applied force.  In our model the duration of kinetochore-MT attachments depends heavily on the flexural rigidity and intrinsic curvature of the PFs. In particular, our model suggests that a variation of the intrinsic curvature $\varphi$ can cause a switch between stable/unstable attachments of depolymerizing PFs from kinetochores. This seems reasonable since there is a broad distribution of bending angles found in experiments \cite{mcintosh-MTang}. Although the intrinsic curvature of a PF in isolation would be solely determined by the angle and stiffness between GTP/GDP-tubulin blocks, it is not unreasonable to assume that other factors influence the mechanical properties of protofilaments and so regulates the stability of the kinetochore-MT interface.

Our model, therefore, naturally suggests an additional path to vary the type of the attachment from stable to unstable and vice-versa. This mechanical process can of course be seen as complementary to any other accompanying process, like for example the stability regulation by Ndc80 \cite{deluca2006} and Aurora B \cite{cimini2006} or the switch of the MT state from growing to shrinking \cite{akiyoshi}. In particular, the breakdown of the catch-bond could also occur according to two additional pathways: (i)
the dissociation of fibril loops and (ii) the fracture of MT-PFs. A possible hypothesis is that Aurora B is responsible for the dissociation of fibril loops as it phosphorylates the terminal ends of the Ndc80 complex \cite{deluca2006}. To our best knowledge the influence of chemical or biological agents on the elastic properties of MTs and PFs and their tips has not been studied and it will be exciting to see whether future experiments prove our predictions.

\subsection*{Comparison with other models}

Earlier models for the kinetochore-microtubule interface make use of the curling of PFs to show how force can be produced by the geometry of the protofilament \cite{molodtsov2,efremov,shtylla}. These studies focus mainly on the load-bearing capabilities of the PFs and do not go into detail on the mechanism of the coupling to the kinetochore. While our work is building on these insights, we include some very recent experimental results into our model allowing us to explore the kinetochore interface in more detail, and make some specific predictions based on them.

In contrast to the biased-diffusion based Hill sleeve \cite{hill-sleeve,joghunt} or the motor-protein based \cite{civek} models, our model does not investigate the biochemical aspects of the kinetochore-MT interface, but focus instead on its mechanics. The biochemical properties of the interface that are needed, i.e. the fibril-tip attachment strength to neighbor fibril tips and the MT surface, are not readily available and have to be fit. We furthermore investigate slightly different phenomena than solely chromosome transport and breathing for which these models were designed for. Our model sheds light instead on the genesis of the kinetochore-MT interface together with the role of the geometrical conformation of the PF tips.

An approach that is similar in spirit to ours was recently published in Ref. \cite{zaytsev2013}. There the authors also model the collective action of kinetochore fibrils on the MTs, but the methodology employed is quite different. While in that paper the authors explore the kinetochore-MT interface in a mean-field approximation using rate equations, we simulate directly the movement of each fibril and its interactions with the MT separately and then look at the emergent behavior. Tellingly, in both studies the cooperation of kinetochore fibrils is found to be essential to strengthen and stabilize the interface. Another difference, of course, is that we explicitly make use of the geometrical details of the protofilament curl and base on it predictions on the emergence of the kinetochore-microtubule interface.

Our results fit neatly between different facets of the study of the kinetochore-microtubule interface and condense established experimental results into a new kind of model. Using molecular and stochastic dynamics simulations, we are able to elucidate and predict some very new aspects of the attachment of kinetochore fibrils to microtubule protofilaments.

\subsection*{Conclusions}

In this paper we introduce a computational model of the kinetochore-microtubule interface based on the synthesis of different experimental observations showing that fibrils extend from the kinetochore and capture MT tips \cite{mcintosh-fibrils,mcintosh-ktmt}, kinetochore proteins form entangled networks \cite{dong-ktmt}, the essential kinetochore protein complex Ndc80 interacts in different ways with curved and straight protofilaments \cite{alushin2010} and Ndc80 complexes bind to each other \cite{ciferri2008}. 
Hence, we assume that individual kinetochore fibrils are composed by different essential protein-complexes and can be treated as building blocks for the kinetochore-MT interface. We then employ a three-dimensional MT model treating the tubulin hetero-dimers as wedge shaped blocks that can have two configuration corresponding to GTP and GDP-bound tubulin. The tubulin blocks organize into protofilaments with intrinsic curvature (GDP-tubulin) or without (GTP-tubulin). The kinetochore fibrils are modelled as bead-spring polymers that can attach to the top-side of straight protofilaments. The fibril tips also attach to each other forming effective loops into which curling protofilament-tips can hook. The model provides a clear illustration on how the conformation of MT-PFs influence
the stability of their attachments with the kinetochore that is quite general and robust.
Beside its application to the experiments reported in Ref. \cite{akiyoshi}, our model represents a general computational tool that could be useful to guide future experimental investigations of MTs-kinetochore interactions.

\section*{Acknowledgments}
We thank Zoe Budrikis and Alessandro Sellerio for useful discussions and comments.
ZB and SZ are supported by the European Research Council through the Advanced Grant 2011 SIZEFFECTS. Work in the laboratory of HM is funded by FEDER through the Operational Competitiveness Programme  -- COMPETE and by National Funds through FCT  - Funda{\,c}{\~a}o para a Ci{\^e}ncia e a Tecnologia under the project FCOMP-01-0124-FEDER-015941 (PTDC/SAU-ONC/112917/2009),
the Human Frontier Science Program and the 7th framework program grant PRECISE from the European Research Council. 


\section*{Supplementary Movie Captions}

\paragraph*{Movie S1. Model of the Microtubule - Kinetochore Interface} 
Kinetochore molecules are assumed to form uniform units represented as fibrils, here shown in red. The tips of the fibrils, dark red, can sense and attach to the outer surface of an incident microtubule (blue) if the MT straight, while fibril-tips do not attach directly to MT-tips that curl. The kinetochore fibril-tips also attach to each other when brought into close contact, a circumstance that is especially likely when the fibrils attach to the MT as neighbours. When fibril tips attach to each other, loops are formed into which curling protofilaments of a depolymerising MT can hook and thus exert force on the kinetochore.

\paragraph*{Movie S2. Detail of the kinetochore - MT interface with a single protofilament.} 
This movie shows a single protofilament interacting with a bunch of kinetochore fibrils from two different angles. The incident straight protofilament captures some kinetochore fibril-tips. When hydrolysed and curling, the protofilament-tip hooks into the loops formed by the fibrils and exerts a force - the loop is visibly pulled out of the mass of the other fibrils, while the fibril bases, as well as the protofilament base, are held static and are not allowed to move. Depolymerisation of the protofilament increases the force exerted on the kinetochore - the fibril loop stretches further - until the interface fractures.

\paragraph*{Movie S3. Fracture of the kinetochore - MT interface I: straight protofilament.}
This movie shows the interface of a straight PF with -- for simplicity -- only one attached kinetochore fibril. After initial thermalization, the fibril base is moved with constant velocity, resulting in the unwrapping of the fibril during which the pulling force is not transmitted onto the kinetochore-protofilament interface. After the unwrapping is complete, the fibril is pulled off the PF, after enough force built up to overcome the potential attaching the fibril tip to the PF.

\paragraph*{Movie S4. Fracture of the kinetochore - MT interface II: curling protofilament.}
We use the same setup as in movie S3, with the difference that the bases of the kinetochore fibrils are not held fixed, but each base is attached to every other by linear springs, to emulate the kinetochore body. At first no external force acts on the fibrils and the curling of the protofilament transports the fibril bundle a short distance until external forces are switched. Then the kinetochore fibril bases are pulled with constant velocity which results in the uncurling of the depolymerizing protofilament and eventual detachment.

\paragraph*{ Movie S5. Fracture of the kinetochore - MT interface III: curling protofilament, fibril-bond breakdown.} We use a similar setup as in movie S4, with the following differences: we reduce the number of fibrils to two and the binding energy between the fibril tips by 50\%. When the kinetochore fibril bases are pulled with constant velocity, the MT starts to uncurl slightly, but then the force on the fibril tips becomes too high and they detach from each other.

\clearpage

\section*{Supplementary Figures}

\makeatletter 
\renewcommand{\thefigure}{S\@arabic\c@figure} 
\makeatother
\setcounter{figure}{0}

\begin{figure*}[ht]
\centering
\includegraphics[width=0.8\textwidth]{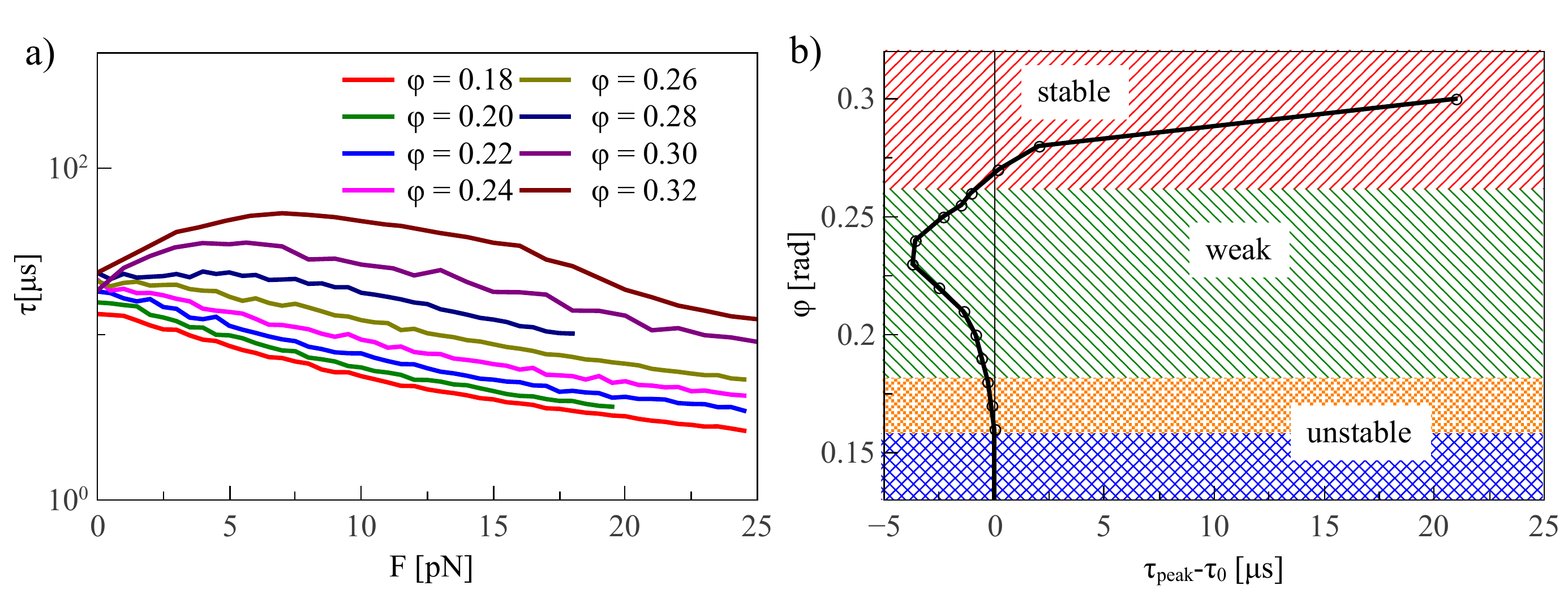}
\caption{{ Kinetochore-MT attachment destabilization.} a) The detachment time as a function of the pulling force 
for different values of the protofilament angle $\varphi$. The peak only occurs for large  $\varphi$ and disappears
for smaller $\varphi$. b)  The stability of the catch-bond behaviour depends strongly on $\varphi$. The PF could switch between stable and unstable attachments by changing $\varphi$.
}
\label{fig:3}
\end{figure*}

\begin{figure*}[ht]
\centering
\includegraphics[width=0.5\textwidth]{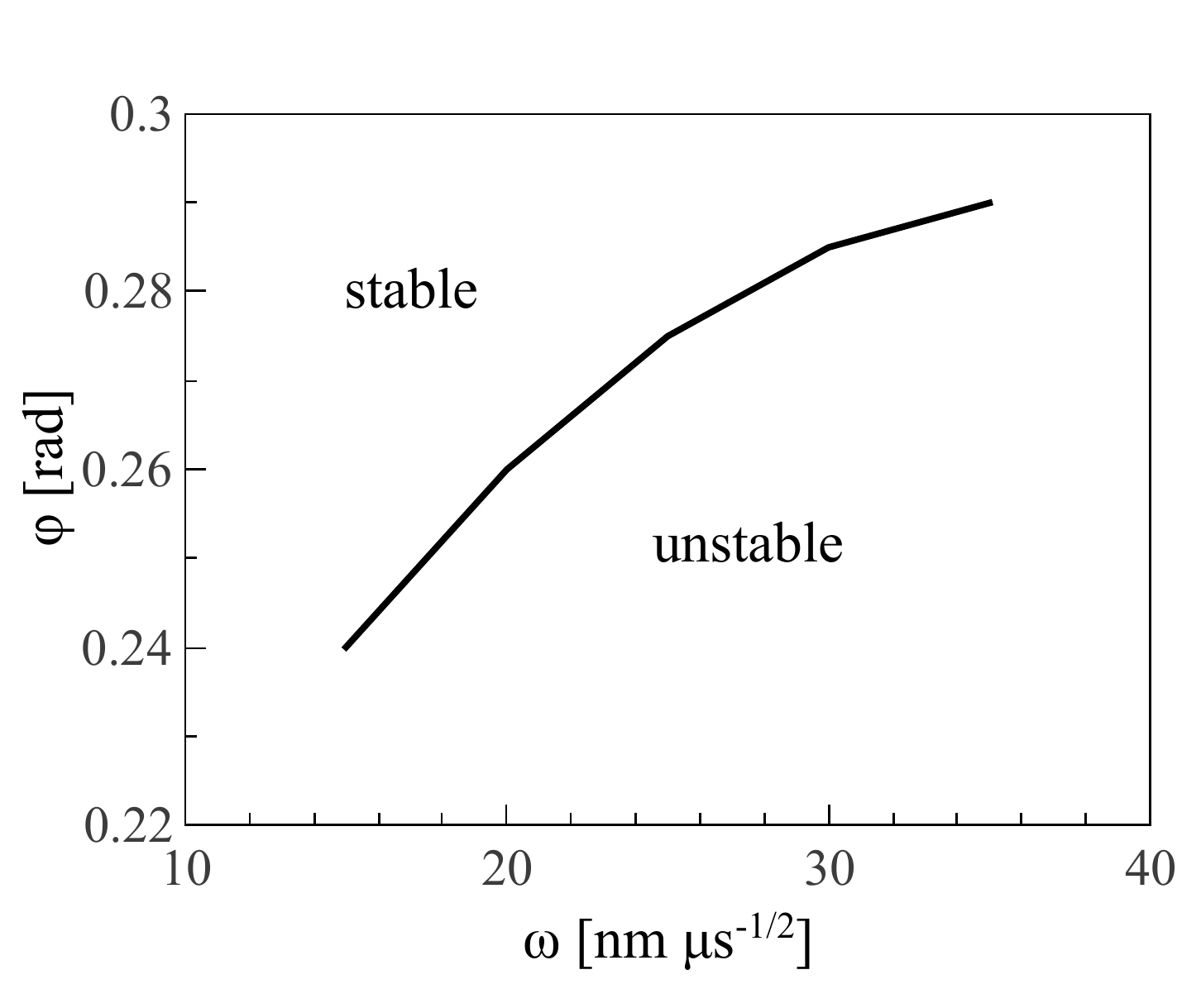}
\caption{{ Phase diagram.} The value of the destabilization angle depends on the diffusion parameter $\omega$, thus
defining a phase diagram for the catch-bond stability for a given flexural rigidity $B$.  
}
\label{fig:4}
\end{figure*}

\end{document}